# The effect of travel restrictions on the spread of a highly contagious disease in Sweden

Martin Camitz[1,2] & Fredrik Liljeros[3,4]

[1]Theoretical Biological Physics, Department of Physics, Royal Institute of Technology, Stockholm, [2]Swedish Institute for Infectious Disease Control, Solna, [3]Department of Medical Epidemiology and Biostatistics, Karolinska Institute, Solna, [4]Department of Sociology, Stockholm University, Stockholm, Sweden

___________________________________________________________________

**Abstract**

Travel restrictions may reduce the spread of a contagious disease that threatens public health. In this study we investigate what effect different levels of travel restrictions may have on the speed and geographical spread of an outbreak of a disease similar to SARS. We use a stochastic simulation model of the Swedish population, calibrated with survey data of travel patterns between municipalities in Sweden collected over three years. We find that a ban on journeys longer than 50 km drastically reduces the speed and the geographical spread of outbreaks, even with when compliance is less than 100%. The result is found to be robust for different rates of inter-municipality transmission intensities. Travel restrictions may therefore be an effective way to mitigate the effect of a future outbreak.

___________________________________________________________________

**1. Introduction**

Knowledge about the speed at which a contagious disease travels between geographical regions, is vital for making decisions about the most effective intervention strategies. The actual routes a disease will take are highly determined by how individuals travel in regions and between regions [1-4]. As was shown during the SARS outbreak [5], the travel patterns of today enable contagious diseases to spread to far corners of the globe at alarming rates. This exposes the need for a new type of model that incorporates travel networks. Recently, Hufnagel et al. [6] demonstrated how a simple stochastic model in conjunction with data on aviation traffic could be used to simulate the global spread of the SARS epidemic. Using a stochastic transmission model on both a city level and globally, with each city interconnected by the international aviation network, they produced results in surprising agreement with reports of the actual case.

This study applies a modified version of the Hufnagel model to Sweden to predict the effect that travel restrictions may have on the geographical spread of an outbreak. Instead of using only the aviation network, which connects only some 30 towns in Sweden, we use survey data on *all* inter-municipal travel. Our choice of a stochastic modeling approach [7] is based on the fact that it acts out the highly random initial phase of an epidemic better than does the traditional deterministic approach[8, 9]. The article is organized as follows: we first present the survey data used to estimate travel intensities between different municipalities in Sweden. We then introduce the simulation model to simulate the spread of the diseases and to study the effect of travel restrictions. Following this, we present the results of the simulations. We conclude our study with a discussion of the validity of the model and possible conclusions for future policy interventions.

**2. Data and Methods**

For this study, we use data from a random survey carried out by Statistics Sweden from 1999 through 2001. A total of 17,000 individuals took part in the survey, constituting 71.9% of the selection. 34,816 distinct inter-municipal trips were reported [10]. An inter-municipal journey is defined between two points where the individual lives, works, or conducts an errand. In other words, we treat a journey between home and work as several trips if the traveler makes stops on the way for errands, provided that a municipal border is crossed between each stop. The data was weighted to correspond to one day and to the entire population for ages 6 to 84. As it turned out, roughly 1% of the data was erroneous in a way that was not negligible and was consequently removed[*]. From this set, we estimate a *travel intensity matrix*, with each element corresponding to the one-way travel intensity between two municipalities. The number of populated elements was 11,611 (to be compared with the size of the matrix, $289 \times 289 = 83,521$). Even though the matrix gives a good picture of the traveling pattern in Sweden, we must treat any intensity between two specific communities with care. This is true especially for small communities with only a single or very few journeys made between them.

A total of nine scenarios are simulated with 1000 realizations each, to study the effects of three levels of travel restrictions as a control measure, for three different levels of the global inter-community infectiousness parameter, $\gamma$, which in Hufnagel's case is used to calibrate the model. 60 days was chosen as the simulation period as this gives plenty of time for a possible extinction to occur and for all stochasticity to play out its part in all but the smallest and most distant municipalities. They each start with a single infectious individual in Stockholm and treat the country as an island isolated from inflow of disease and with no possibility of traveling abroad. The traveling restrictions are divided into the following levels. In the first level, we use the complete intensity matrix. In the following two, we have removed data corresponding to journeys longer than 50 km and journeys longer than 20 km, respectively. The simulations are henceforth designated SIM, SIM50, and SIM20. In Figure 1, the data sets are displayed in geographical plots.

---

[*]The erroneous records were long distance journeys, mostly between odd communities in unreasonably short time. Had they not been removed, their influence would have been significant, accelerating the spread across the country. The correct data was irretrievable but the effect of its absence is deemed within the margin of error for long distance journeys.

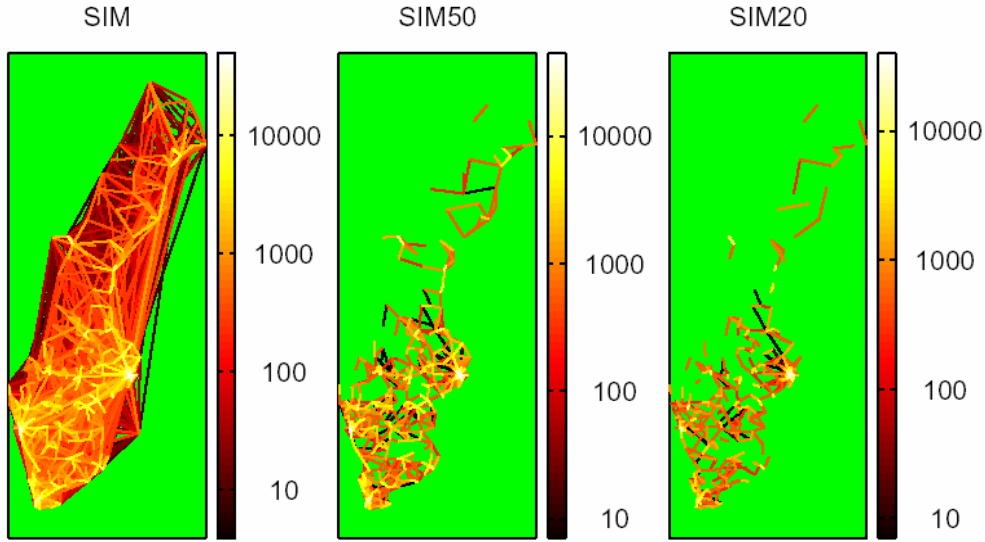

Figure 1: The inter-municipal travel network with travel intensities indicated by color lines. The scale is logarithmic in trips per day. SIM shows the complete data set. In SIM50 and SIM20, all journeys longer than 50 km and 20 km, respectively, have been removed. The lines are drawn between the population centers of each municipality, so in many cases the trips are shorter than the lines representing them.

As a complement we consider the case that the travel restrictions are not obeyed wholly by the public. Perhaps 5% do not head the restrictions resulting in a small but non-zero intensity for trips longer than the set restrictions. Full 1000-run simulations are made at varying levels of distance restrictions and compliance, resulting in a mesh surface of the incidence.

We use a simplified version of the model suggested by Hufnagel et al. [6]. The individuals in both models can be in four different states:

- **S** *Susceptible.*
- **L** *Latent, meaning infected but not infectious.*
- **I** *Infectious*
- **R** *Recovered and/or immune.*

The rate at which individuals move from one category to the next is governed by the intensity parameters: the attack rate $\alpha = 0.55$, $\beta = 0.21$ which is the inverse infectious time and $\nu = 0.19$, the inverse latency time [11, 12]. The transitions between different states in a municipality $i$ can be viewed schematically as follows:

$$S_i + I_i \xrightarrow{\alpha} L_i + I_i, \quad L_i \xrightarrow{\nu} I_i, \quad I_i \xrightarrow{\beta} R_i. \quad (1)$$

To the first of these, we further assume a contribution from every other municipality $j$, resulting in the following possible transitions:

$$S_i + I_j \xrightarrow{\gamma_{j,i}} L_i + I_j. \qquad (2)$$

As the process is assumed to be Markovian as in Hufnagel, the time between two events, $\Delta t$, is random, taken from an exponential distribution,

$$\Delta t \in \mathrm{Exp}(1/Q) \qquad (3)$$

where $Q$ is the total intensity, the sum of all independent transmission rates:

$$Q_i^L = \alpha I_i \frac{S_i}{N_i} + \sum_{j \neq i} \gamma_{j,i} I_j \frac{S_i}{N_i},$$
$$Q_i^I = \nu L_i,$$
$$Q_i^R = \beta I_i.$$

$$Q = \sum_i Q_i^L + \sum_i Q_i^I + \sum_i Q_i^R. \qquad (5)$$

The component $\gamma_{j,i}$ (note the reversed indexes) is the inter-municipal infectiousness corresponding to the one-way route $j$ to $i$. If $\omega_{j,i} = M_{j,i} / \sum_i M_{j,i}$, where $M_{j,i}$ is the travel intensity, i.e. $\omega_{j,i}$ is the probability that a traveler in $j$ will choose the route $j$ to $i$, then $\gamma_{j,i} = \gamma \omega_{j,i}$. In case restrictions are active, this expression is further scaled row wise to match the smaller mass of the matrix. $\gamma$ is the global inter-municipal infectiousness parameter mentioned above. We use the approximate estimated by Hufnagel based on data from the actual outbreak, $\gamma = 0.27$. The parameter $\gamma$ is influenced by the total travel intensity, the medium of travel and as we've seen, the propensity for travel in different communities. We would like to calibrate our model in a similar way, but as we have no outbreak data for Sweden, we need to see whether changes in $\gamma$ drastically alter our conclusions. To get an idea of its effect, we compare Hufnagel's estimate of $\gamma$ with other possible values. As $\gamma$ is an infectiousness parameter on the same standing as $\alpha$, we argue that $\alpha = 0.55$ must be an upper bound for $\gamma$. To get a lower bound we extrapolate linearly from these, producing $0.13$.

Although this is not mentioned in the Hufnagel's original work, the expression above means that everybody, regardless of where they live, is equally prone to travel outside their home municipality. This is a heavy assumption indeed as it depends on the function of the municipality as a suburb or self-sufficient community varies, as it does for airports across the globe for various other reasons. One of the strengths of Hufnagel's model is that it seems to be forgiving towards many simplifications, this one included, with the correct choice of g. We investigated corrections for this assumption, such as row wise scaling according to the known probability for travel, but found little effect on absolute incidence and none on the qualitative conclusions of the current study. As such, we were reluctant to stray from Hufnagel's model.

The simulation runs as follows: First we move forward in time with a random step $\Delta t$ given by Expression 3. We then select the event that will occur with a probability proportional to the corresponding intensity. All intensities are updated according to the new state, and the process is repeated until the disease dies out or the simulation period, 60 days, is passed.

## 3. Results

The results for all nine scenarios are plotted geographically and color-coded according to the mean incidence in Figure 2.

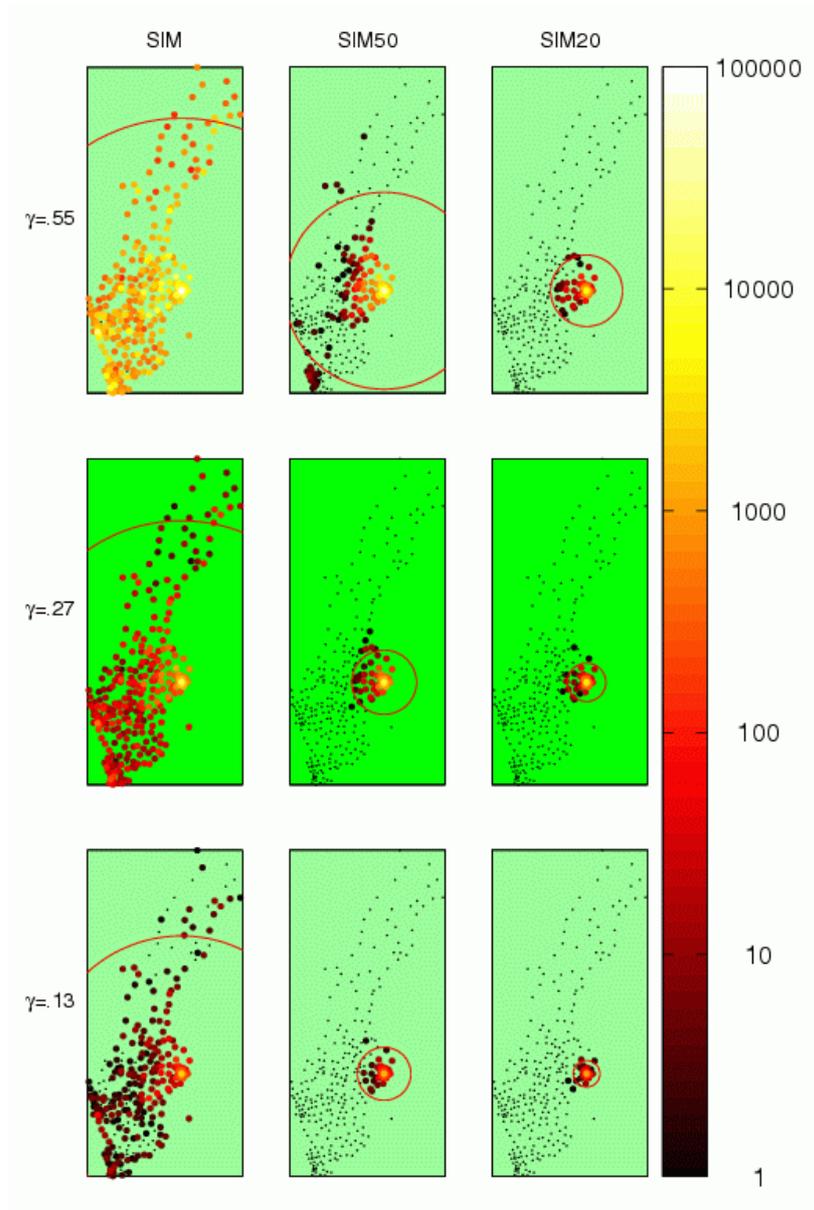

Figure 2: Geographical plot of the municipalities logarithmically color-coded according to the mean incidence after 60 days. SIM depicts the complete data set. In SIM50 and SIM20, all journeys longer than 50 km and 20km, respectively, have been removed. The red circle signifies the mean extent of the epidemic from Stockholm.

A scenario with no restrictions results in an outbreak in which a majority of the municipalities become affected regardless of $\gamma$. Only the incidence differs. A ban on journeys longer than 50 kilometers stifles the dynamics of the outbreak. For the two lower values of $\gamma$, we see that the disease remains in the Stockholm area after 60 days, and for the high value of $\gamma$, the disease has not managed to spread far from the densely populated areas around the largest Swedish cities. Prohibiting journeys longer than 20 kilometers will result in an even slower spread with a small number of afflicted municipalities, mainly localized around Stockholm. What is more, the total incidence after 60 days as well as the incidence in each municipality drops as we impose the restrictions.

Table 1 compares the country's total incidence in the three simulations for which Hufnagel's estimate of $\gamma$ was used. Table 2 presents the incidence broken down into a few selected municipalities.

Table 1: The table shows the main results along with miscellaneous information about the simulation. Figures refer to simulated values at the end of the run, 60 days or earlier in case of extinction. The mean when applicable is taken over the complete set of 1000 realizations. By *incidence* we mean the number of infectious. *Inter-municipal infections* is the percentage of the total number of infected that caught the disease via inter-municipal infection. There are 289 municipalities in Sweden and the population is approximately 8.9 million.

|  | SIM | | | SIM50 | | | SIM20 | | |
|---|---|---|---|---|---|---|---|---|---|
|  |  | Min | Max |  | Min | Max |  | Min | Max |
| Total number of infected | 242 562 | 1 | 1 613 506 | 113 107 | 1 | 711 901 | 44 758 | 1 | 328 196 |
|     Percentage of population | 2.7% | 0% | 18% | 1.3% | 0% | 8.0% | .05% | 0% | 3.7% |
|     Inter-municipal infections | 25.9% | 0% | 66.7% | 22.0% | 0% | 66.7% | 15.9% | 0% | 66.7% |
| Incidence | 58 391 | 0 | 393 117 | 27 131 | 0 | 172 580 | 10 607 | 0 | 77 985 |
|     Percentage of population | 0.66% | 0% | 4.4% | 0.31% | 0% | 1.95% | 0.12% | 0% | 0.88% |
| Afflicted municipalities | 196.7 | 1 | 289 | 34.6 | 1 | 74 | 23.7 | 1 | 47 |
| Mean incidence in municipalities | 202.0 | 0 | 1 360 | 93.9 | 0 | 597 | 36.7 | 0 | 270 |
| Mean time for extinction (days) | 2.9 | 0.0 | 41.0 | 3.5 | 0.0 | 31.9 | 3.6 | 0.0 | 60.3 |
| Mean travel distance (km) | 65.9 | - |  | 22.2 | - |  | 11.3 | - |  |
| Total intensity (millions/day) | 4.3 | - |  | 2.9 | - |  | 1.5 | - |  |
| Inter-municipal one-way routes | 11 694 | - |  | 1 386 | - |  | 797 | - |  |
| Extinction runs | 249 | - |  | 268 | - |  | 305 | - |  |
| Mean time for extinction (days) | 2.9 | 0.0 | 41.0 | 3.5 | 0.0 | 31.9 | 3.6 | 0.0 | 60.3 |
| Afflicted municipalities before extinction | 1.27 | 1 | 4 | 1.29 | 1 | 5 | 1.28 | 1 | 4 |
| Total number of realizations | 1 000 | - |  | 1 000 | - |  | 1 000 | - |  |

The reason for the decrease in incidence is of course the limited transmission paths available to the disease. The disease, after having spread from one municipality to another will constantly be transmitted back into the originating municipality -- provided that there is a flow of travelers in the opposite direction in the travel intensity matrix. Travel restrictions limit both spread to other municipalities and reintroduction. For comparison, if traffic is removed altogether, the incidence in Stockholm will be 917.

The high number of extinction runs may be surprising, but it is in accordance with the theory of Markov processes, which dictates that $1/R_0 = 37\%$ of the realizations should terminate in extinction [13]. In the simulations of our study, a lower value is expected due to spread to and retransmission from other municipalities.

Table 2: A selection of municipalities[†] with the mean incidence, maximum and minimum.

| Municipality | SIM | | | SIM50 | | | SIM20 | | |
|---|---|---|---|---|---|---|---|---|---|
| | Mean | Min | Max | Mean | Min | Max | Mean | Min | Max |
| Stockholm | 13843 | 1 | 82066 | 9685 | 1 | 61577 | 4196 | 1 | 31674 |
| Göteborg | 604.4 | 0 | 25439 | 0.0 | 0 | 0 | 0.0 | 0 | 0 |
| Malmö | 303.3 | 0 | 11052 | 0.0 | 0 | 0 | 0.0 | 0 | 0 |
| Huddinge | 2597 | 0 | 13294 | 1908 | 0 | 10715 | 848 | 0 | 5669 |
| Upplands-Bro | 421.8 | 0 | 2895 | 265.1 | 0 | 2011 | 58.6 | 0 | 979 |
| Norrtälje | 694.6 | 0 | 4646 | 157.1 | 0 | 2173 | 26.0 | 0 | 908 |
| Södertälje | 846.9 | 0 | 5954 | 467.2 | 0 | 4277 | 42.3 | 0 | 2583 |
| Västerås | 635.7 | 0 | 7007 | 19.8 | 0 | 780 | 2.0 | 0 | 277 |
| Eskilstuna | 498.5 | 0 | 4307 | 44.5 | 0 | 1327 | 18.1 | 0 | 890 |
| Umeå | 91.8 | 0 | 4856 | 0.0 | 0 | 0 | 0.0 | 0 | 0 |
| Luleå | 178.1 | 0 | 8209 | 0.0 | 0 | 0 | 0.0 | 0 | 0 |
| Örebro | 424.0 | 0 | 8363 | 0.2 | 0 | 33 | 0.0 | 0 | 0 |
| Jönköping | 166.2 | 0 | 2932 | 0.0 | 0 | 0 | 0.0 | 0 | 0 |
| Linköping | 380.3 | 0 | 4454 | 1.3 | 0 | 108 | 0.0 | 0 | 0 |
| Helsingborg | 138.7 | 0 | 5011 | 0.0 | 0 | 0 | 0.0 | 0 | 0 |
| Borås | 119.2 | 0 | 3036 | 0.0 | 0 | 0 | 0.0 | 0 | 0 |
| Gävle | 449.6 | 0 | 6566 | 16.1 | 0 | 581 | 1.3 | 0 | 108 |
| Ljungby | 24.6 | 0 | 1301 | 0.0 | 0 | 0 | 0.0 | 0 | 0 |
| Hofors | 58.3 | 0 | 1059 | 2.1 | 0 | 252 | 0.0 | 0 | 2 |
| Örkelljunga | 4.1 | 0 | 178 | 0.0 | 0 | 0 | 0.0 | 0 | 0 |

It is also clear how travel restrictions increasingly protect cities the further they are from the capital, the focal point of the infection. The major cities of Gothenburg and Malmö are protected even though traffic into these cities is heavy. In fact, the farthest the disease ever makes it in SIM50 is Ljungby, 1471 km from Stockholm and still some 200 km from Malmö. For SIM20 the farthest city is Uddevalla, 441 km away and a suburb to Gothenburg. The mean reach of the epidemic is only 276 km and 34 km, respectively.

An objection as to the applicability of this model is that in all probability, complete enforcement of the restrictions may not be achievable or even desirable in the case of high priority professionals with crucial functions in society during a crisis situation. Incidence does indeed climb the more restrictions are ignored, but not to such an extent as to render the travel restrictions dubious as a means of disease control, see figure 3. A plot with unrestricted travel, duplicated from figure 2, is given for comparison.. Figure 4 shows a finer spaced mesh of incidence versus restriction distance and compliance. Bear in mind that there was no attempt to correlate the randomness between the simulation sets. Therefore the random numbers used in each is completely independent giving rise to considerable simulation noise. Even though the landscape is rough the trend in both dimensions is clearly visible. Looser travel restrictions and lower compliance means higher incidence.

---

[†] After Stockholm, Gothenburg and Malmö are the largest cities in Sweden. The single most traveled route is that between Stockholm and neighboring Huddinge, traveled by approximately 37,000 peopledaily, each way. The decline in incidence closely follows that in Stockholm. Upplands-Bro is representative of an outer suburb to Stockholm. Södertälje and Norrtälje are nearby towns but are not considered suburbs. Västerås and Eskilstuna are more distant, but have a fair number of commuters. The last four are small towns in southern Sweden, and the remaining ones, Örebro through Luleå, are larger towns at some distance from Stockholm with no notable commuter traffic.

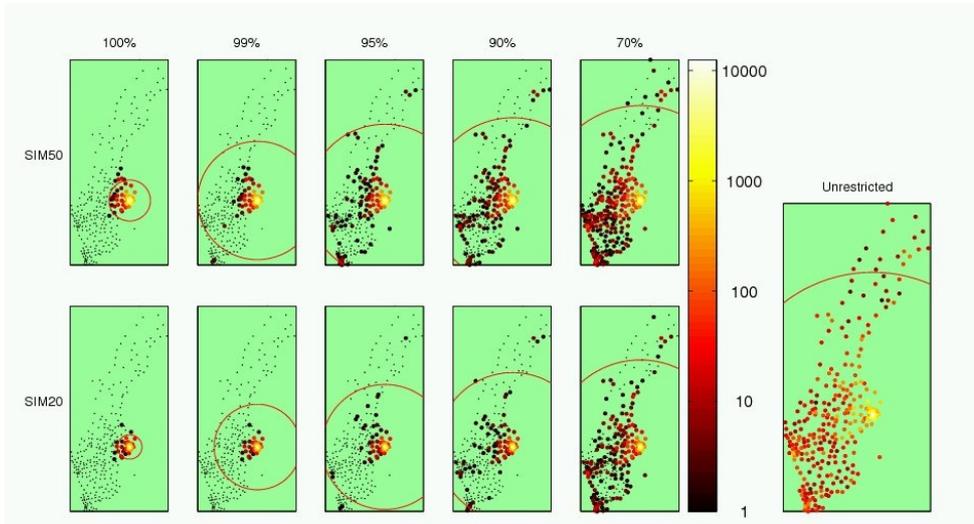

Figure 3: Geographical distribution of the incidence after 60 days shown for SIM50 and SIM20 for different levels of compliance. The left most plot shows the unrestricted case with Hufnagel's original γ-value for comparison. This plot reflects the same data as the one on the middle row, right column of figure 2 but with scale to match the current figure.

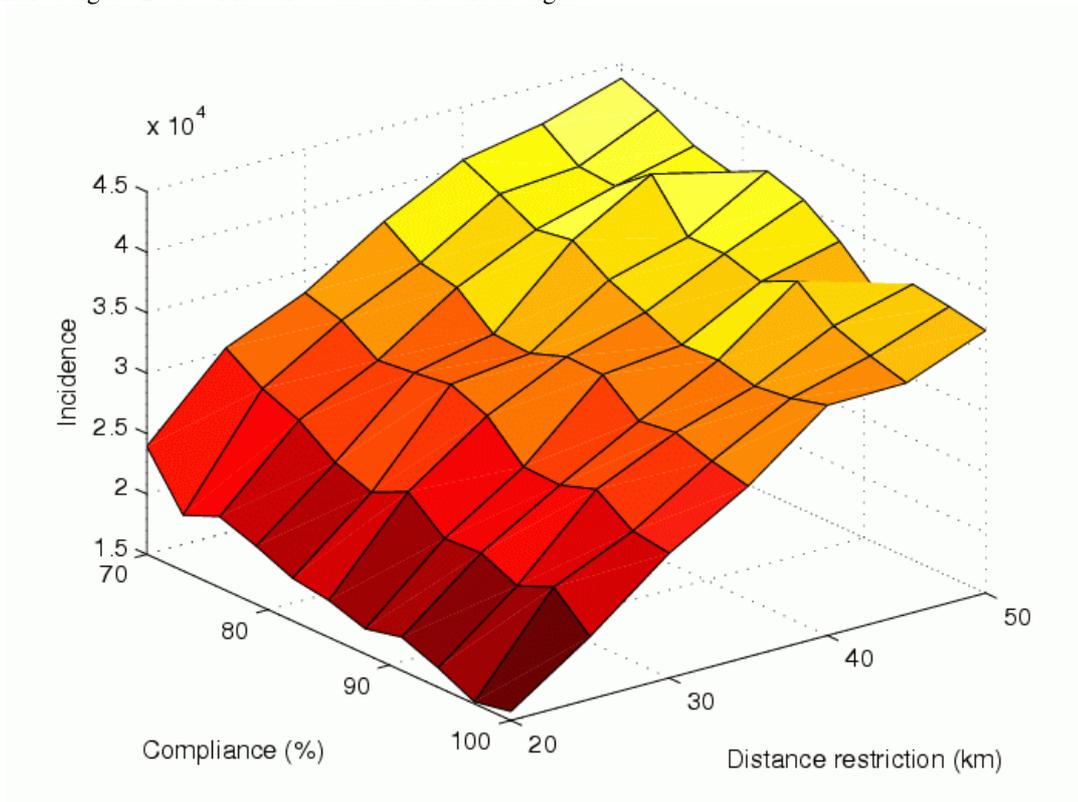

Figure 4: A surface plot showing incidence after 60 days with parameters compliance and distance restrictions on the data axes. 1000 realizations are made for each point. In contrast to other results presented in this paper, extinction runs have been filtered from this plot in order to reduce contributions from simulation noise. The surface has its highest values at high set distance limit and low compliance. Its low values are found at opposite corner.

**4. Discussion**

Our results show clearly that traveling restrictions will have a significant beneficial effect, both reducing the geographical spread and the total and local incidence. This holds true for all three levels of inter-community infectiousness simulated, $\gamma$. $\gamma$ is influenced by many factors, most notably by total travel intensity, but also by the medium of travel, the behavior of the traveler, the model of dispersal by travel and by the infectiousness of the disease. Hufnagel calibrated $\gamma$ using data from the actual outbreak. As mentioned, no attempt was made on our part to find the "true" value of $\gamma$ in the new settings, as no such outbreak data is available for Sweden. This would be considered a flaw for a quantitative study on a SARS outbreak in Sweden. By simulating for different values of the parameter, however, we can be confident in the qualitative conclusion, namely, that the same general behavior can be expected in the unrestricted scenario and in response to the control measures, regardless of $\gamma$.

In light of the fact that inter-municipal travel heavily influences incidence even at a local level, one may justifiably be concerned about the boundary conditions. We treat Sweden as an isolated country, but quite obviously, the incidence will be underestimated for areas with frequent traffic across the borders. This includes in particular the Öresund region around Malmö, and to a lesser extent, international airports and the small towns bordering on Norway and Finland.

Even though there is presently no treatment or vaccine for SARS, results show that limited quarantine as suggested here drastically decreases the risk of transmission and this may well turn out to be the most expedient form of intervention. In many countries, Sweden included, limiting freedom of travel is unconstitutional and must take the form of general recommendations. Additionally, certain professions of crucial importance to society during a crisis situation must be exempt from travel restrictions. The study shows that even if a substantial fraction of the population breaks the restrictions, this strategy is still viable. For other types of disease for which preventive treatment (pandemic flu) or vaccine (small-pox) are available, our results show that long-distance travelers are an important group for targeted control measures.

**Acknowledgements**

This study was supported by The Swedish Institute for Infectious Diseases Control, Swedish Council for Working Life and Social Research, and the European Union Research NEST Project (DYSONET 012911). The authors would like to express their gratitude to Tom Britton and Åke Svensson, Institute of Mathematics,

Stockhholm University and Alden Klovdahl, Social Sciences, Australian National University. The authors are members of S-GEM, Stockholm Group for Epidemic Modeling, www.s-gem.se.